\documentclass[12pt]{article}
\usepackage{epsfig,amsfonts,amssymb}
\usepackage{hyperref}
\usepackage{cite}
\input epsf.sty
\usepackage{cite}

\newcommand{\hab}{}

\def\ZZZ{{\hbox{ Z\kern-1.6mm Z}}}
\def\RRR{{\hbox{ R\kern-2.4mm R}}}
\def\CCC{{\hbox{ C\kern-2.0mm C}}}
\def\zzz{{\hbox{z\kern-1mm z}}}

\newcommand{\ten}{{(10)}}
\newcommand{\bet}{{( b )}}

\newcommand{\qq}{k}
\newcommand{\pp}{l}
\newcommand{\nn}{\nonumber \\}

\newcommand{\vt}{\vartheta}

\newcommand{\vtau} {\vec \tau}
\newcommand{\vj} {\vec J}
\newcommand{\vxi} {\vec \xi}
\newcommand{\vu} {\vec u}
\newcommand{\htau} {\vec \eta}
\newcommand{\vc}{\vec\chi}
\newcommand{\vpsi} {\vec \psi}

\newcommand{\qeq}{{\hbox{=\kern-2.3mm ? \kern.5mm }}}
\renewcommand{\qeq}{=}

\newcommand{\rrho}{r}
\newcommand{\bA}{{\bf A}}
\newcommand{\tx}{\wt x}
\newcommand{\bG}{{\bf G}}
\newcommand{\bF}{{\bar F}}
\newcommand{\bbb}{{\bar b}}
\newcommand{\gam}{\tau}
\newcommand{\eps}{\epsilon}
\newcommand{\vareps}{\varepsilon}
\newcommand{\ra}{\rangle}
\newcommand{\la}{\langle}
\newcommand{\T}{\chi_{T}(k)}
\newcommand{\Tm}{\chi_{T}(k')}
\newcommand{\Cn}{{\cal C}_n}
\newcommand{\vp}{\varphi}
\newcommand{\ve}{\varepsilon}
\newcommand{\tl}{\lambda}
\newcommand{\dt}{(\vec \nabla T)^2}
\newcommand{\hp}{{\wh\Phi}}
\newcommand{\hq}{{\wh Q_B}}
\newcommand{\he}{{\wh\eta_0}}
\newcommand{\ha}{{\wh{A}}}
\newcommand{\lllb}{\Bigl\langle\Bigl\langle}
\newcommand{\rrrb}{\Bigr\rangle\Bigr\rangle}
\newcommand{\tf}{\wt f}
\newcommand{\sss}{{\cal L}_{av}}
\newcommand{\bx}{\bar x}
\newcommand{\bw}{\bar w}
\newcommand{\ws}{{\wt\sigma}}
\newcommand{\wrh}{{\wt\rho}}
\newcommand{\wv}{{\wt v}}

\newcommand{\vv} {\bar v}
\newcommand{\uu} {\bar u}

\newcommand{\K}{{\rm K_1}}
\newcommand{\Kt}{{\rm \widetilde K_1}}

\newcommand{\B}{b'}
\newcommand{\C}{c\,'}
\newcommand{\bB}{\bar b'}
\newcommand{\Bu}{B_{\vec u}}
\newcommand{\VV}{{\cal V}}
\newcommand{\BB}{{\cal B}}
\newcommand{\DD}{{\cal D}}
\newcommand{\BBB}{{\cal B}}
\newcommand{\II}{{\cal I}}
\newcommand{\AAA}{{\cal A}}
\newcommand{\GG}{{\cal G}}
\newcommand{\KK}{{\cal K}}
\newcommand{\fff}{{\bf f}}
\newcommand{\ccc}{{\bf c}}
\newcommand{\FF}{{\cal F}}
\newcommand{\JJ}{{\cal J}}
\newcommand{\HH}{{\cal H}}
\newcommand{\MM}{{\cal M}}
\newcommand{\CC}{{\cal C}}
\newcommand{\bC}{{\bf C}}
\newcommand{\OO}{{\cal O}}
\newcommand{\QQ}{{\cal Q}}
\newcommand{\PP}{{\cal V}}
\newcommand{\EE}{{\cal E}}
\newcommand{\LL}{{\cal L}}
 \newcommand{\rrr}{\rangle\rangle}
\newcommand{\half}{{1\over 2}}
\newcommand{\wt}{\widetilde}
\newcommand{\wh}{\widehat}
\newcommand{\wc}{\wt}
\newcommand{\wb}{\bar}
\newcommand{\RR}{{\cal R}}
\newcommand{\NN}{{\cal N}}
\newcommand{\TT}{{\cal T}}
\newcommand{\bg}{\bar g}
\newcommand{\ba}{\bar a}
\newcommand{\bc}{\bar c}
\newcommand{\bd}{\bar d}
\newcommand{\bb}{\bar b}
\newcommand{\bT}{\bar \Theta}
\newcommand{\SSS}{{\cal S}}
\newcommand{\tlx}{\left(\tilde \lambda ; X^0(0) \right)}
\newcommand{\al}{\alpha}

\newcommand{\tk}{\tilde \kappa}

\newcommand{\ppp}{\prime\prime}

\newcommand{\omk}{\omega_n(\vec k)}
\newcommand{\onk}{\omega^{(N)}_{\vec k_\perp}}
\newcommand{\tI}{\wt\II}
\newcommand{\hI}{\wh\II}
\newcommand{\nI}{\II}

\newcommand{\cp}{\check\Phi}
\newcommand{\cps}{\Psi}
\newcommand{\crh}{\check\rho}
\newcommand{\cs}{\check\sigma}
\newcommand{\cv}{\check v}
\newcommand{\com}{\check\Omega}

\newcommand{\be}{\begin{equation}}
\newcommand{\ee}{\end{equation}}
\newcommand{\ben}{\begin{eqnarray}\displaystyle}
\newcommand{\een}{\end{eqnarray}}

\newcommand{\refb}[1]{(\ref{#1})}
\newcommand{\p}{\partial}
\newcommand{\sectiono}[1]{\section{#1}\setcounter{equation}{0}}
\newcommand{\subsectiono}[1]{\subsection{#1}\setcounter{equation}{0}}

\newcommand{\zet}{\zeta}

\newcommand{\gsim}{\stackrel{>}{\sim}}
\newcommand{\lsim}{\stackrel{<}{\sim}}

\newcommand{\Lamb}{\Lambda}

\def\one{{\hbox{ 1\kern-.8mm l}}}
\def\zero{{\hbox{ 0\kern-1.5mm 0}}}

\def\wa{{\wh a}}
\def\wb{{\wh b}}
\def\wc{{\wh c}}
\def\wc{\check}
\def\wdd{{\wh d}}

\newcommand{\bi}{{\bf i}}

\renewcommand{\theequation}{\thesection.\arabic{equation}}

\newcommand{\bea}[1]{\begin{eqnarray}\label{#1} }
\newcommand{\eea}{\end{eqnarray}}

\newcommand{\wJ}{\wt J}
\newcommand{\bN}{{\bf N}}

\newcommand{\aaa}{b}

\newcommand{\eqref}{\refb}

\newcommand{\un}{{\rm u}}

\usepackage{bm}
\usepackage[table]{xcolor}
\def\rpnote#1{{\color{magenta} #1}}
\def\arnote#1{{\color{blue} #1}}
\def\asnote#1{{\color{red} #1}}

\def\figb{
\def\JPicScale{0.8}
\ifx\JPicScale\undefined\def\JPicScale{1}\fi
\unitlength \JPicScale mm

}

\def\figg{
\def\JPicScale{0.8}
\ifx\JPicScale\undefined\def\JPicScale{1}\fi
\unitlength \JPicScale mm
\begin{picture}(120,60)(0,0)
\linethickness{1.15mm}
\put(20,50){\line(1,0){100}}
\linethickness{0.75mm}
\multiput(30,60)(1.33,-1.33){8}{\multiput(0,0)(0.11,-0.11){6}{\line(1,0){0.11}}}
\linethickness{0.75mm}
\multiput(40,50)(1.33,1.33){8}{\multiput(0,0)(0.11,0.11){6}{\line(1,0){0.11}}}
\linethickness{0.75mm}
\multiput(60,60)(1.33,-1.33){8}{\multiput(0,0)(0.11,-0.11){6}{\line(1,0){0.11}}}
\linethickness{0.75mm}
\multiput(70,50)(1.33,1.33){8}{\multiput(0,0)(0.11,0.11){6}{\line(1,0){0.11}}}
\linethickness{0.75mm}
\multiput(90,60)(1.33,-1.33){8}{\multiput(0,0)(0.11,-0.11){6}{\line(1,0){0.11}}}
\linethickness{0.75mm}
\multiput(100,50)(1.33,1.33){8}{\multiput(0,0)(0.11,0.11){6}{\line(1,0){0.11}}}
\put(25,45){\makebox(0,0)[cc]{$k$}}

\put(50,45){\makebox(0,0)[cc]{$k$}}

\put(85,45){\makebox(0,0)[cc]{$k$}}

\put(110,45){\makebox(0,0)[cc]{$k$}}

\end{picture}

}

\begin{document}

\begin{flushright}
DAMTP-2014-27\\
HRI/ST1406
\end{flushright}

\vskip 12pt

\baselineskip 24pt

\begin{center}
{\Large \bf  String Perturbation Theory Around Dynamically Shifted Vacuum}

\end{center}

\vskip .6cm
\medskip

\vspace*{4.0ex}

\baselineskip=18pt

\centerline{\large \rm Roji Pius$^a$, Arnab Rudra$^b$ and Ashoke Sen$^a$}

\vspace*{4.0ex}

\centerline{\large \it ~$^a$Harish-Chandra Research Institute}
\centerline{\large \it  Chhatnag Road, Jhusi,
Allahabad 211019, India}
\centerline{\large \it ~$^b$Department of Applied Mathematics and Theoretical Physics}
\centerline{\large \it Wilberforce Road, Cambridge CB3 0WA, UK}

\vspace*{1.0ex}
\centerline{\small E-mail:  rojipius@mri.ernet.in, A.Rudra@damtp.cam.ac.uk, sen@mri.ernet.in}

\vspace*{5.0ex}

\centerline{\bf Abstract} \bigskip

In some string theories, e.g. SO(32) heterotic string theory on Calabi-Yau manifolds, 
a massless field with a tree level potential can acquire a tachyonic
mass at the one loop level, forcing us to quantize the theory around a new background that
is not a solution to the classical equations of motion and hence is not described by a
conformally invariant world-sheet theory. We describe a systematic procedure for carrying
out string perturbation theory around such backgrounds.

\vfill \eject

\baselineskip=18pt

\baselineskip=18pt

\tableofcontents

\sectiono{Introduction}  \label{s1}

In many $\NN=1$ supersymmetric compactification of string theory down to 3+1 dimensions,
we have $U(1)$ gauge fields with Fayet-Iliopoulos (FI) terms generated at one 
loop\cite{DSW,ADS,DIS,greenseiberg} (see \cite{1304.2832,1404.5346} 
for a recent perspective on this). 
By choosing suitable linear combination of these gauge fields we can ensure that only one
gauge field has FI term.
Typically there are also massless scalars $\phi_i$
charged under this U(1) gauge field. If $q_i$ is the charge carried by $\phi_i$ then the
presence of the FI term generates a term in the potential of the form
\be \label{e1}
{1\over g^2}\, \left( \sum_i q_i \phi_i^* \phi_i - C\, g^2\right)^2
\ee
where $C$ is a numerical constant that determines the coefficient of the FI term and
$g$ is the string coupling.
$C$ could be positive or negative and $q_i$'s for different fields could have different signs.
As a result when we expand the potential in powers of $\phi_i$ around the perturbative
vacuum $\phi_i=0$, some of these scalars can become tachyonic.\footnote{It was shown in
\cite{ADS} that for any compactification of SO(32) heterotic string theory preserving (2,2)
world-sheet supersymmetry, there is always at least one such tachyonic scalar, leading to the
existence of a stable supersymmetric vacuum.}
It is clear from the form
of the effective potential that the correct procedure to compute physical quantities is to shift
the corresponding fields so that we have a new vacuum where
$\sum_i q_i \langle\phi_i^*\rangle \langle\phi_i\rangle = C\, g^2$, and quantize string theory
around this new background. However since classically the $C\, g^2$ term is absent from
this potential \refb{e1}, this new vacuum is not a solution to the
{\it classical} equations of motion. As a result on-shell 
methods\cite{1209.5461,Belopolsky,dp,Witten,1304.7798}, which require that we begin
with a conformally invariant world-sheet theory, is not suitable for carrying out a systematic
perturbation expansion around this new background.

Although the above example provides the motivation for our analysis, we shall address this
in a more general context. At the same time we shall simplify our analysis by assuming that
only one scalar field is involved instead of multiple scalar fields. So we consider a general
situation in string theory where at tree level we have a massless real scalar with a non-zero
four point coupling represented by a potential
\be \label{e2}
A \, \phi^4 + \cdots
\ee
where $\cdots$ denote higher order terms.
We suppose further that at one loop the scalar receives a negative contribution 
$-2\, C g^2$
to its 
mass$^2$. Here $A$ and $C$ are $g$-independent constants. 
Then the total potential will be
\be \label{e3}
A\, \phi^4 - C\, g^2 \, \phi^2 + \cdots \, .
\ee
This has a minimum at
\be 
\phi^2 = {1\over 2} {C\over A}\, g^2 + \cdots\, .
\ee
Our goal will be to understand how to systematically develop string perturbation theory
around this new background and also to correct the expectation value of $\phi$ due to
higher order corrections.
If we had an underlyng string field theory that is fully consistent at the quantum level, {\it e.g.}
the one described in \cite{sft}, then that would provide a natural framework for addressing
this issue. Our method does not require the existence of an underlying string field theory,
although the requirement of gluing compatibility of the local coordinate system that we shall 
use is borrowed from string field theory.

The method we shall describe can be used to address
other similar problems in string theory where loop correction induces small shift in the vev
of a massless field. For example suppose we have a massless field $\chi$ with a tree level cubic
potential and suppose further that one loop correction generates a tadpole for this field.
Then from the effective field theory approach it is clear that there is a nearby perturbative
vacuum where the field $\chi$ is non-tachyonic. Usual string perturbation theory does not tell
us how to deal with this situation, but the method we describe below can be used in this case
as well.

There are of course also problems involving tadpoles of massless fields without tree
level potential, {\it e.g.} of the kind discussed 
in \cite{FS} and many follow up papers.
As of now our method does not offer any new insight into such problems.

The rest of the paper is organised as follows. In \S\ref{s2} we describe the procedure for
constructing amplitudes in the presence of a small shift in the vacuum expectation value of
a massless scalar following the procedure of \cite{earlyref}. 
We also discuss systematic procedure for determining the
shift by requiring absence of tadpoles. In \S\ref{s3} we show that  the amount of shift in the scalar,
needed to cancel the tadpole, depends on the choice of local coordinate system that we use
to construct the amplitudes.
However 
general physical amplitudes
in the presence of the shift are independent of the
choice of local coordinate system as long as we use a gluing compatible system
of local coordinates for  defining the amplitude. This is our main
result. For simplicity we restrict our analysis in \S\ref{s2} and \S\ref{s3} to bosonic string theory,
but in \S\ref{super} we discuss generalization of our analysis to include NS sector
states in heterotic and superstring theories. In \S\ref{sinfra} we describe the procedure for
regulating the spurious infrared divergences in loops,
arising from the fact that the shift in the vacuum
renders some of the originally massless states massive.

Earlier attempts to analyze string theory in a shifted background can be found in
\cite{0410101}.

\sectiono{Systematic construction of the new vacuum}  \label{s2}

We shall carry out our analysis under 
several simplifying assumptions. These are made mainly
to keep the analysis simple, but we believe that none of these (except 
\ref{no4}) is necessary.
\begin{enumerate}
\item We shall assume that there is a 
symmetry under which $\phi\to -\phi$ so that amplitudes with odd number of external $\phi$
fields vanish. 
\item We shall assume that $\phi$ does not mix with any other physical or unphysical states of
mass level zero even when quantum corrections are included.
\item Shifting the $\phi$ field can sometimes induce tadpoles in other massless fields.
If there is a tree level potential for this field then we can cancel the tadpole by 
giving a vacuum expectation value (vev) to that field and determine the
required vev by following the same procedure that we used to determine the shift in $\phi$.
We shall assume that such a situation does not arise and that
$\phi$ is the only field that needs to be shifted.
However extension of
our analysis to this more general case should be straightforward.
\item \label{no4} If on the other hand shifting $\phi$ leads to the tadpole of a massless field
which
has vanishing tree level potential then it is not in general 
possible to find a nearby vacuum where all tadpoles vanish. In this case  
the vacuum is perturbatively
unstable. We shall assume that this is not the case here. 

\item When the theory has other massless
fields besides $\phi$ but their tadpoles vanish, 
then the situation can be dealt with in the manner discussed in 
\S7.2 of \cite{1209.5461} and will not be discussed here any further.\footnote{In 
the special case of the D-term potential in supersymmetric
theories discussed in \S\ref{s1} the dilaton tadpole does not vanish in the perturbative
vacuum\cite{DSW,AtickS,1209.5461,1403.5494}, but
is expected to vanish in the shifted vacuum since the latter
has zero energy density.}
\item In  this section and in \S\ref{s3} we shall restrict our analysis
to the bosonic string theory. However the result can be generalized to include 
the case where $\phi$
is Neveu-Schwarz (NS) sector field in the heterotic string theory or NS-NS sector field in
type IIA or IIB string theory. This is discussed briefly in \S\ref{super}.
\end{enumerate}

As discussed in detail in \cite{1311.1257,1401.7014}, for computing 
renormalized masses and S-matrix of general string
states we need to work with off-shell string theory. This requires choosing a set of gluing compatible
local coordinate system on the (super-)Riemann surfaces. The result for off-shell amplitude 
depends on the choice of local coordinates, but the renormalized masses and S-matrix elements
computed from it are independent of this choice. 
Our analysis will be carried out in this context.

The off-shell amplitudes do not directly compute the off-shell Green's functions. Instead they
compute truncated off-shell Green's functions. If we denote by $G^{(n)}(k_1, b_1; \cdots k_n,b_n)$
the $n$-point off-shell Green's function of fields carrying quantum numbers $\{b_i\}$ and
momenta $\{k_i\}$, then the truncated off-shell Green's functions are defined as
\be \label{eoff}
\Gamma^{(n)}(k_1, b_1; \cdots k_n,b_n) = G^{(n)}(k_1, b_1; \cdots k_n,b_n)\, 
\prod_{i=1}^n (k_i^2 + m_{b_i}^2)\, ,
\ee
where $m_{b}$ is the {\it tree level} mass of the state carrying quantum number $b$.
The usual on-shell amplitudes of string theory compute $\Gamma^{(n)}$ at $k_i^2=-m_{b_i}^2$.
This differs from the S-matrix elements by  multiplicative wave-function renormalization factors
for each external state and also due to the fact that the S-matrix elements require replacing 
$m_{b_i}^2$'s by physical mass$^2$'s in this formula. However from the knowledge of
off-shell amplitude $\Gamma^{(n)}$ defined in \refb{eoff} we can extract the physical 
S-matrix elements following the procedure described in \cite{1311.1257,1401.7014}.

Our goal is to study what happens when we switch on a vev of $\phi$.
For this we shall first consider a slightly different situation. Suppose that $\phi$ is an 
exactly marginal deformation in string theory and furthermore that it remains marginal even
under string loop corrections. In this case there is no potential for $\phi$ and we can give any
vacuum expectation value $\lambda$ to $\phi$. The effect of this is to deform the world-sheet
action by a term $\lambda \int d^2 z \, V_\phi(z,\bar z)$ where $V_\phi$ is the vertex operator of
the zero momentum $\phi$ state. In the string amplitude, obtained by integrating the correlation
functions of the underlying conformal field theory (CFT) on moduli spaces of punctured Riemann
surfaces, this introduces a term
\be 
\exp[\lambda \int \, d^2 z V_\phi(z,\bar z)] 
=\sum_{m=0}^\infty {\lambda^m\over m!} \, \left( \int\, d^2 z V_\phi(z,\bar z)\right)^m\, .
\ee
The effect of the $\left( \int\, d^2 z V_\phi(z,\bar z)\right)^m$ 
term is to convert $\Gamma^{(n)}$ to $\Gamma^{(n+m)}$
with $m$ insertions of zero momentum $\phi$ state. This if we denote by $\Gamma^{(n)}_\lambda$
the deformed off-shell amplitudes then we have the relation
\be \label{edefrule}
\Gamma^{(n)}_\lambda (k_1, b_1; \cdots k_n,b_n) = \sum_{m=0}^\infty {\lambda^m\over m!}\, 
\Gamma^{(n+m)}(k_1, b_1; \cdots k_n,b_n;  0, \phi; \cdots  0, \phi)\, ,
\ee
where we have denoted the quantum number $b$ labelling the field $\phi$ by $\phi$ itself.
There are altogether $m$ insertions of $0,\phi$ in the argument of $\Gamma^{(n+m)}$
on the right hand side. In any expression of this kind that we shall be using later, the number
of insertions of $(0,\phi)$ can be figured out by subtracting from the superscript 
of $\Gamma$ the number of explicit $(k_i, b_i)$ factors in the argument of $\Gamma$.
The insertion of $(0,\phi)$ factors in \refb{edefrule} has to be interpreted as the result of
taking the zero momentum limit of a general amplitude where the external $\phi$ states
carry non-zero and different momenta. As we shall see, individual contributions to the
right hand side have tadpole divergence in the zero momentum limit. We shall discuss ways to
regulate this later.
A direct proof of \refb{edefrule} in a quantum field theory has been given in appendix
\ref{sa}.

Now in our case the field $\phi$ has a potential even at the tree level and hence does not
represent an exactly marginal deformation. If we nevertheless go ahead and try to define
a deformed theory using \refb{edefrule}, we encounter the following problem. We have
\be \label{e4}
\Gamma^{(1)}_\lambda (0,\phi) =  \sum_{m=0\atop m\, odd}^\infty {\lambda^m\over m!}\, 
\Gamma^{(1+m)}(0, \phi; 0, \phi; \cdots  0, \phi) \, ,
\ee
where we have used the postulated $\phi\to -\phi$ symmetry to restrict the sum over
$m$ to odd values only.
It will be useful to express the right hand side as sum over contributions from different
genera. Thus we write\footnote{We have dropped an overall $1/g^2$ factor from the
definition of $\Gamma^{(n)}$ so that we can drop a $g^2$ factor from the definition of
the propagator later in \refb{edefdelta}. If we use the standard convention where $1/g^2$
appears as an overall multiplicative factor in the
tree level action, then the propagator $\Delta$ will have an
extra factor of $g^2$ and the $\Gamma^{(n)}$'s will have extra factor of $1/g^2$. If we denote
these by $\Gamma_{standard}=g^{-2} \Gamma$ and $\Delta_{standard}=g^2\Delta$, then
it is straightforward to check that all our subsequent equations hold with $\Gamma$ replaced
by $\Gamma_{standard}$ and $\Delta$ replaced by $\Delta_{standard}$ without any extra
factor of $g^2$.}
\be \label{e4a}
\Gamma^{(1)}_\lambda (0,\phi) =  \sum_{s=0}^\infty  \, 
g^{2s}\sum_{m=0\atop m\, odd, \, m+2s \ge 3}^\infty 
{\lambda^m\over m!} \, \Gamma^{(1+m;s)}(0, \phi; 0, \phi; \cdots   0, \phi)\, ,
\ee
where $\Gamma^{n;s}$ denotes genus $s$ contribution to the $n$-point function in the
unperturbed theory.
The $s=0$, $m=3$ term on the right hand side is non-zero since it is proportional to the 
four point function of zero momentum $\phi$ states and is proportional to $A$ according
to \refb{e2}. Thus in the deformed theory there is a zero momentum $\phi$ tadpole
proportional to $\lambda^3$. This is
clearly not an acceptable vacuum at the tree level
since it will give divergent results for higher point amplitudes.

But now consider the effect of one loop correction given by the $m=1$, $s=1$ term
on the right hand side of \refb{e4a}. 
This is non-zero and represents the second term in \refb{e3}. 
We now see that by a suitable choice of
$\lambda$, given by the solution to
\be \label{eleading}
{1\over 6}\, \lambda^3  \, \Gamma^{(4;0)}(0, \phi; 0, \phi; 0,\phi;  0, \phi)
+ \lambda \, g^{2} \, \Gamma^{(2;1)}(0, \phi; 0, \phi)=0\, 
\ee
 we can cancel the net contribution to the
$\phi$ tadpole to order $g^3$. This vanishes for three distinct values of $\lambda$, one of which
is given by $\lambda=0$ and the other two are related by $\phi\to -\phi$ symmetry.
We shall be considering the situation where 
in  the $\lambda=0$ vacuum the field $\phi$ is tachyonic and hence this
solution needs to be avoided.
This fixes $\lambda$ to a specific value of order $g$ up to the $\phi\to -\phi$ symmetry. 

With this choice of $\lambda$
we make $\Gamma^{(1)}_\lambda (0,\phi)$ vanish to order $g^3$.
To extend the analysis to higher order in $g$ we express the condition of vanishing of
\refb{e4a} as
\ben \label{eiter}
&&{1\over 6}\, \lambda^2  \, \Gamma^{(4;0)}(0, \phi; 0, \phi; 0,\phi;  0, \phi)
+ g^{2} \, \Gamma^{(2;1)}(0, \phi; 0, \phi) \nonumber \\
&=& - \sum_{m,s=0\atop m\, odd; \, m+2s\ge 5}^\infty {1\over m!} \, \lambda^{m-1}\, 
g^{2s} \, \Gamma^{(1+m;s)}(0, \phi; 0, \phi; \cdots   0, \phi)\, ,
\een
and then solve this equation iteratively using the leading order solution \refb{eleading}
as the starting point. 
Note that in arriving at \refb{eiter} we have divided \refb{e4a} by $\lambda$, thereby removing
the trivial solution $\lambda=0$.
To each order in iteration we substitute on the right hand side of
\refb{eiter} the solution for $\lambda$ to the previous order and then solve 
\refb{eiter}. Due to the $\phi\to -\phi$ symmetry and the fact that the genus expansion
is in powers of $g^2$, the solution for $\lambda$ takes the form
\be \label{elambdaexpan}
\lambda^2 =\sum_{n=0}^\infty A_n g^{2n+2}\, ,
\ee
for constants $A_n$. Furthermore note that by adjusting $\lambda^2$ to order $g^{2n+2}$,
we can satisfy \refb{eiter} to terms of order $g^{2n+2}$, \i.e.\ make the right hand side
of \refb{e4a} vanish to order $g^{2n+3}$.\footnote{Due to the
$\phi\to -\phi$ symmetry and the fact that the genus expansion is in powers of $g^2$,
the  contributions to $\Gamma^{(1)}_\lambda (0,\phi)$ involve only odd powers of $g$
after using \refb{elambdaexpan}.
Thus making $\Gamma^{(1)}_\lambda (0,\phi)$ vanish to order $g^{2n+3}$ also makes
it vanish to order $g^{2n+4}$.}

There are however some additional subtleties in this analysis, since 
the individual terms on the
right hand side of \refb{eiter} can diverge due to $\phi$ tadpole. These divergences arise from
regions of moduli integral where a Riemann surface degenerates into two distinct Riemann
surfaces connected by a long handle. As mentioned at the end of \S\ref{s1},
we shall proceed with the assumption that the only relevant divergences are associated
with tadpoles of $\phi$. To deal with these divergences we need to first regularize  
these divergences, solve for $\lambda$, and
at the end remove the regulator. 
For this we shall work with a choice of gluing compatible
local coordinates according to the procedure described in \S3.2  of
\cite{1401.7014} and express a general
amplitude contributing to the right hand side of \refb{e4} 
as a sum of products of `one particle irreducible' (1PI) amplitudes joined by 
the propagator
\be \label{edefdelta}
\Delta = {1\over 4\pi}\,
\int_0^\infty ds \, \int_0^{2\pi} d\theta \, e^{-(s-i\theta)L_0 - (s+i\theta) \bar L_0}\, .
\ee
The divergence in individual contributions come from the $s\to\infty$ limit on the
right hand side of \refb{edefdelta}. We shall regulate the divergence by replacing
$\Delta$ by
\be \label{edeltareg}
{1\over 4\pi}\,
\int_0^\Lambda ds \, \int_0^{2\pi} d\theta \, e^{-(s-i\theta)L_0 - (s+i\theta) \bar L_0}\, ,
\ee
where $\Lambda$ is a fixed large number. 
The relevant divergence in $\Delta$ comes from the choice where the propagating
state is a zero momentum $\phi$. Since $L_0$ and $\bar L_0$ vanish for zero momentum
$\phi$, the contribution to $\Delta$
from this term
goes as $P_\phi \, \Lambda/2$, where $P_\phi$ 
denotes projection to the CFT state corresponding to zero momentum
$\phi$. 
It will be convenient to define,\footnote{For computation of
$\Gamma^{(1)}_\lambda$ using \refb{e4a} all propagators
carry zero momentum and hence the second line of \refb{edefbardelta} is irrelevent, but this
definition will be useful for analyzing $\Gamma^{(n)}_\lambda$ in the next section.}
\ben \label{edefbardelta}
&& \Delta_\phi = {1\over 2} \Lambda P_\phi, \quad \bar\Delta = \Delta - \Delta_\phi \, \quad
\hbox{for momentum $k=0$}\, , \nonumber \\
&& \bar\Delta = \Delta \quad \hbox{for $k\ne 0$}\, .
\een
We furthermore denote by $\bar \Gamma$ the contribution to an amplitude
obtained by taking sum of products of 1PI contributions joined by the modified
propagator $\bar \Delta$. 
Thus the difference between the full amplitude $\Gamma$ and the modified amplitude
$\bar \Gamma$ is controlled by $\Delta_\phi$.
We expect the individual contributions to $\bar \Gamma$ to be
free from divergence associated with zero momentum propagator
in the $\Lambda\to\infty$ limit, since the contribution to
$\bar \Delta$ from a massive
state of mass $m$ is given by $(1 - e^{-\Lambda m^2/2})/m^2$, 
while the contribution from the other massless states
are expected to vanish due to the assumed vanishing of the corresponding tadpoles.
{}From this argument it is also clear that the $\Lambda$ dependence of 
$\bar \Gamma$ will
come through exponentially suppressed terms for large but finite $\Lambda$.
The full amplitude is obtained as sum of products of $\bar\Gamma$'s joined by
$\Delta_\phi$'s. 

\begin{figure}
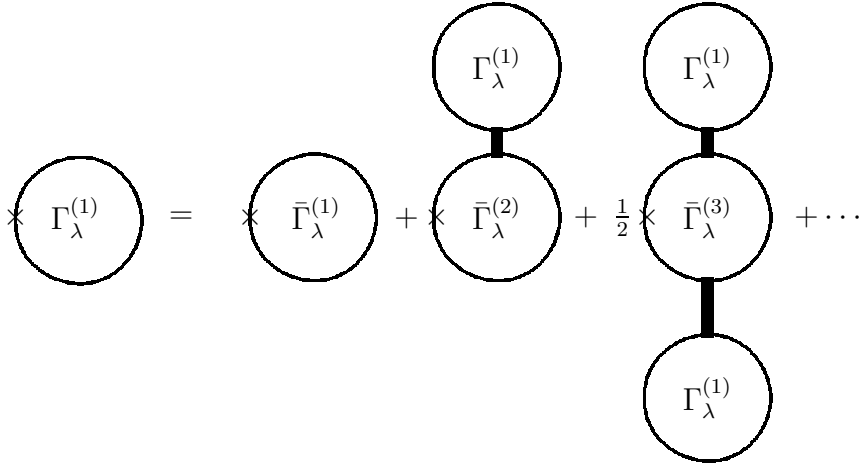

\figb
\caption{Pictorial representation of \refb{e6}. The $\times$ denotes the vertex
operator associated with the external $\phi$ state. The $\phi$ vertex operators carrying
factors of $\lambda$ are not displayed explicitly. The thick line denotes the $\phi$
propagator
$\Delta_\phi$.} \label{fb}
\end{figure}

We now define $\bar\Gamma^{(n)}_\lambda$ as in \refb{edefrule} 
with all the $\Gamma$'s replaced
by $\bar\Gamma$'s on the right hand side.
This allows us to
express
$\Gamma^{(1)}_\lambda(0,\phi)$
as (see Fig.~\ref{fb})
\be \label{e6}
\Gamma^{(1)}_\lambda(0,\phi) = \sum_{k=0}^\infty \, {1\over k!}\,
\bar\Gamma^{(1+k)}_{\lambda} (0,\phi; 0, \phi; \cdots ;0, \phi)
\, \left(\Delta_{\phi} \Gamma^{(1)}_\lambda(0, \phi)\right)^k\, .
\ee
By repeated use of \refb{e6} we can express $\Gamma^{(1)}_\lambda(0,\phi)$ as sum of
products of $\bar\Gamma_\lambda$ factors and $\Delta_\phi$'s, but we shall not write down the
explicit formula.

Now suppose that $\lambda$ to order $g^{2n+1}$, obtained by solving \refb{eiter} to 
order $g^{2n+2}$,
or equivalently by demanding the vanishing of $\Gamma^{(1)}_\lambda(0,\phi)$  to order
$g^{2n+3}$,
has finite limit as $\Lambda\to\infty$. Our goal will be to prove that the result also holds
with $n$ replaced by $n+1$.
For determining
$\lambda$ to next order, we need to compute
$\Gamma^{(1)}_\lambda(0,\phi)$ to order $g^{2n+5}$ and then require this to
vanish. 
Using the result that
$\bar\Gamma^{(2)}_\lambda(0,\phi;0,\phi)$ has its expansion beginning at order
$g^2$,  one can show that
in order to compute the right hand side of \refb{e6} to order $g^{2n+5}$ we need
to know the 
$\Gamma^{(1)}_\lambda(0, \phi)$ appearing on the { right} 
hand side of \refb{e6} at most to order
$g^{2n+3}$. 
By assumption this contribution vanishes.
Thus only the $k=0$ term contributes on the right hand side of 
\refb{e6}, showing that to order $g^{2n+5}$, $\Gamma^{(1)}_\lambda(0,\phi)
=\bar \Gamma^{(1)}_\lambda(0,\phi)$. 
This means that 
in order to determine the order $g^{2n+3}$ correction to $\lambda$ we can replace 
$\Gamma$ by $\bar\Gamma$ on the right hand side of \refb{eiter}.\footnote{$\Gamma$'s
appearing on the left hand side of \refb{eiter} are in any case equal to the corresponding
$\bar\Gamma$'s.}
Since $\bar \Gamma$'s by construction are finite as $\Lambda\to\infty$
we see that the order $g^{2n+3}$ correction to $\lambda$ is also finite as $\Lambda\to\infty$.
This proves the desired result.

We shall see in \S\ref{s3} that even though $\lambda$ determined using this procedure
is finite in the $\Lambda\to\infty$ limit, it is ambiguous \i.e.\ it depends on the choice of local
coordinate system used for the computation. Nevertheless all physical amplitudes
will turn out to be free from this ambiguity.

\sectiono{General amplitudes at the new vacuum} \label{s3}

Once $\lambda$ is determined, we can use \refb{edefrule} to compute the general
$n$-point amplitude in the deformed theory. However we need to ensure that this has
finite $\Lambda\to\infty$ limit and that it is unambiguous, {\it e.g.} independent of the
choice of local coordinates used to define the 1PI amplitudes.
Our discussion will follow closely
that of \S7.6 of \cite{1209.5461}. However in \cite{1209.5461} the massless
tadpoles were assumed to cancel at every genus while here we consider the case where the
cancelation is between the contributions from different genera. Furthermore in \cite{1209.5461}
the massless fields were assumed to have flat potential while here the relevant field $\phi$ has
a potential even at the tree level.

\begin{figure}
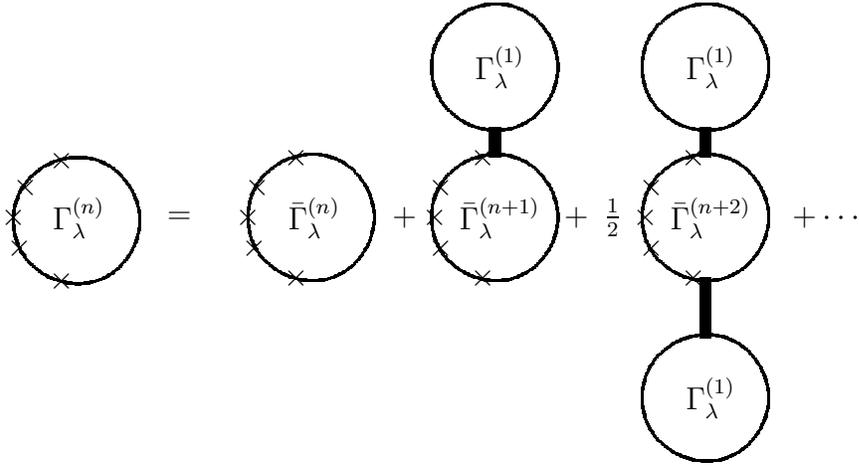


\figc
\caption{Pictorial representation of \refb{e6gen}. The $\times$ denotes the vertex
operator associated with the external states carrying quantum numbers
$(k_1,b_1;\cdots k_n,b_n)$. The $\phi$ vertex operators carrying
factors of $\lambda$ are not displayed explicitly. The thick line denotes the propagator
$\Delta_\phi$.} \label{fc}

\end{figure}

First we examine the issue of finiteness in the $\Lambda\to\infty$ limit. 
We shall assume that all the external states (labelled by $1,\cdots n$ in \refb{edefrule}) carry
generic non-zero momentum.
Thus possible source of zero momentum propagators on the right hand side of
\refb{edefrule} are propagators which connect two Riemann surfaces, one of which  carry all the
external states $1,\cdots n$ and possibly some of the zero momentum $\phi$ vertex 
operators and the other one carries only zero momentum $\phi$ vertex operators. 
For studying the divergences associated with these zero momentum propagators, we
shall define $\bar\Delta$ as in \refb{edefbardelta} and
introduce the amplitudes $\bar \Gamma$ by following
the procedure described below \refb{edefbardelta}.
It is easy to see that we now have the following generalization of \refb{e6} (see Fig.~\ref{fc}) 
\be \label{e6gen}
\Gamma^{(n)}_\lambda(k_1,b_1;\cdots k_n,b_n) = 
\sum_{k=0}^\infty \, {1\over k!}\, 
\bar\Gamma^{(n+k)}_{\lambda} (k_1,b_1;\cdots k_n,b_n; 0, \phi; \cdots 0, \phi)
\left(\Delta_{\phi} \Gamma^{(1)}_\lambda(0,\phi)\right)^k\, .
\ee
Now $\lambda$ has been chosen so that $\Gamma^{(1)}_\lambda(0, \phi)$ vanishes.
This shows that 
only the $k=0$ term contributes to the right hand side of \refb{e6gen}.
Since $\bar\Gamma^{(n)}_{\lambda} (k_1,b_1;\cdots k_n,b_n)$
is finite as $\Lambda\to\infty$,
this establishes that 
$\Gamma^{(n)}_\lambda(k_1,b_1;\cdots k_n,b_n) $ also  has a finite limit as
$\Lambda\to\infty$.

We now have to show that $\Gamma^{(n)}_\lambda(k_1,b_1;\cdots k_n,b_n) $ is
independent of the choice of local coordinate system, except for the expected dependence
associated with the off-shell external states carrying 
momenta $k_1,\cdots k_n$. These latter
dependences can be analyzed and treated in the same
way as in \cite{1311.1257,1401.7014}, and we shall not discuss them any further.
To focus on the real issue we can for example concentrate on the case where these
external states are massless states that do not suffer from mass renormalization so that the
corresponding vertex operators do not introduce any dependence on the choice of local
coordinates.\footnote{The wave-function renormalization factors do depend on the choice
of local coordinates, but this can be treated as in \cite{1209.5461}.}
The problematic dependence on local coordinates arises from the following 
source\cite{1209.5461} (\S7.6).
Let us consider two Riemann surfaces $A$ and $B$, 
glued at their punctures $P_1$ and $P_2$ by 
plumbing fixture procedure:
\be \label{eplumb}
w_1 w_2 = e^{-s+i\theta}
\ee
where $w_1$ and $w_2$ are the local coordinates around the punctures $P_1$ and $P_2$
and $(s,\theta)$ are the same variables which appear in the definition \refb{edefdelta} of the
propagator. Having the cut-off $s\le \Lambda$ then corresponds to requiring
\be \label{ec1}
|w_1 w_2| \ge e^{-\Lambda}\, .
\ee
Now suppose we change the local coordinates to $w_1'$, $w_2'$ related to $w_1$ and $w_2$
via the relations
\be
w_1 = f(w_1'), \quad w_2 =  g(w_2')\, ,
\ee
where $f$ and $g$ are some specific functions satisfying $f(0)=0$, $g(0)=0$. 
Since for small $w_1$ and $w_2$ we have
\be 
w_1 = f'(0) w_1', \quad w_2=g'(0)\, w'_2\, ,
\ee
we can express \refb{ec1} as
\be \label{eshift0}
|w_1'w_2'| \ge e^{-\Lambda'}, \quad \Lambda' = \Lambda + 2\xi_A+2\xi_B,
\quad \xi_A\equiv {1\over 2} \ln |f'(0)|, \quad \xi_B\equiv  {1\over 2} \ln\, |g'(0)|\, .
\ee
Here
$A$ and $B$ refer to the two Riemann surfaces that are connected by the propagator whose
change we are considering.
$A$ and $B$ are abstract symbols
which characterize information on the external legs, genus, as well as the point in the moduli space
we are in, since $f'(0)$ and $g'(0)$ could depend on all these informations. Note however 
that $\xi_A$ does not depend on the Riemann surface $B$ and $\xi_B$ does not
depend on the Riemann surface $A$ {\it provided we choose a gluing compatible local
coordinate system}.

\refb{eshift0} shows that changing the local coordinates correspond to effectively changing the
cut-off. This in turn changes the regulated propagator \refb{edeltareg} by
\be \label{eshift}
\delta_\Lambda \Delta =  ( \xi_A+\xi_B) \, P_\phi\, .
\ee
We have ignored the change in the propagator due to massive states since they are
exponentially suppressed in the $\Lambda\to\infty$ limit, and also from other massless states
with vanishing tadpole since their effect can be taken care of by following the procedure
described in \cite{1209.5461} (see also \cite{catoptric}).
This justifies the appearance of the projection operator
$P_\phi$ to zero momentum $\phi$ states.

Now consider the right hand side of \refb{eiter}. Individual terms in this
expression are divergent in the $\Lambda\to\infty$ limit, but when we use the value of
$\lambda$ by solving \refb{eiter} to certain order, and then substitute this on the right hand
side of \refb{eiter} to compute $\lambda$ to the next order, 
the right hand side of \refb{eiter} has been shown to have a finite $\Lambda\to\infty$
limit. Nevertheless since under a general change of local coordinates different terms on
the right hand side of \refb{eiter} could have their effective $\Lambda$'s changed differently,
there is no guarantee that the right hand side of \refb{eiter} will remain unchanged. In other
words, although we have argued that $\lambda$ determined by solving \refb{eiter} has a
finite $\Lambda\to\infty$ limit, it could depend on the choice of local coordinates. Similar
arguments show that the right hand side of 
\refb{e6gen} could also depend on the choice of local coordinates. Our goal will be to show that
the explicit dependence of the right hand side of \refb{e6gen} on the choice of local coordinates
through the cut-off $\Lambda$ cancels against the implicit dependence of \refb{e6gen}
on the choice of local
coordinates through $\lambda$, so that the final result for
$\Gamma^{(n)}_\lambda(k_1,b_1;\cdots k_n,b_n)$ is independent of the choice of local
coordinates.

{}From now on we shall consider infinitesimal changes in local coordinates so that the
$\xi_A$ and $\xi_B$ in \refb{eshift} are infinitesimal. 
Using \refb{eshift} 
the change in $\Gamma^{(n)}_\lambda(k_1,b_1;\cdots k_n,b_n)$ 
due to a change in $\Lambda$ induced by change of local coordinates take the form
\be \label{eshift3}
\delta_\Lambda
\Gamma^{(n)}_\lambda(k_1,b_1;\cdots k_n,b_n)=\sum_{A,B} 
\Gamma^{(n+1)}_{\lambda,A}(k_1,b_1;\cdots k_n,b_n;0,\phi)\, 
(\xi_A + \xi_B) \Gamma^{(1)}_{\lambda,B}(0,\phi)\, ,
\ee
where sum over $A$ and $B$ denotes sum over all Riemann surfaces (including integration
over moduli and sum over genus) 
that contributes to $\Gamma^{(n+1)}_{\lambda}(k_1,b_1;\cdots k_n,b_n;0,\phi)$
and $\Gamma^{(1)}_{\lambda}(0,\phi)$ respectively. This can be expressed as
\ben \label{efeq}
&& \left\{ \sum_A \xi_A\Gamma^{(n+1)}_{\lambda,A}(k_1,b_1;\cdots k_n,b_n;0,\phi)\, 
\right\} \, \sum_B  \Gamma^{(1)}_{\lambda,B}(0,\phi)
 \nonumber \\ && + 
 \left\{ \sum_A \Gamma^{(n+1)}_{\lambda,A}(k_1,b_1;\cdots k_n,b_n;0,\phi)\, 
\right\} \, \sum_B  \xi_B \Gamma^{(1)}_{\lambda,B}(0,\phi)\, , \nonumber \\
&=& \left\{ \sum_A \xi_A\Gamma^{(n+1)}_{\lambda,A}(k_1,b_1;\cdots k_n,b_n;0,\phi)\, 
\right\} \,  \Gamma^{(1)}_{\lambda}(0,\phi)\, \nonumber \\ &&
+ \Gamma^{(n+1)}_{\lambda}(k_1,b_1;\cdots k_n,b_n;0,\phi)
\, \sum_B  \xi_B\Gamma^{(1)}_{\lambda,B}(0,\phi)\, . 
\een
The first term on the right hand side vanishes since $\lambda$ has been chosen so that
$\Gamma^{(1)}_{\lambda}(0,\phi)$ vanishes. This allows us to write \refb{efeq} as
\be \label{enfin}
\delta_\Lambda
\Gamma^{(n)}_\lambda(k_1,b_1;\cdots k_n,b_n)
= \Gamma^{(n+1)}_{\lambda}(k_1,b_1;\cdots k_n,b_n;0,\phi)
\, \sum_B \xi_B \Gamma^{(1)}_{\lambda,B}(0,\phi)\, . 
\ee
Now suppose $\delta\lambda$ denotes the compensating change in $\lambda$
required to make  $\Gamma^{(1)}_\lambda(0,\phi)$  vanish with the new choice
of local coordinates. Using \refb{edefrule} we see that
this induces a change in $\Gamma^{(n)}_\lambda(k_1,b_1;\cdots k_n,b_n)$ of the form
\be\label{enfin1}
\delta_\lambda
\Gamma^{(n)}_\lambda(k_1,b_1;\cdots k_n,b_n)
= \delta\lambda \, \, \Gamma^{(n+1)}_{\lambda}(k_1,b_1;\cdots k_n,b_n;0,\phi)\, .
\ee
Adding \refb{enfin} to \refb{enfin1} we get the total change in 
$\Gamma^{(n)}_\lambda(k_1,b_1;\cdots k_n,b_n)$:
\be \label{efin}
\delta
\Gamma^{(n)}_\lambda(k_1,b_1;\cdots k_n,b_n)
=  \Gamma^{(n+1)}_{\lambda}(k_1,b_1;\cdots k_n,b_n;0,\phi) 
\left\{ \delta\lambda + \sum_B \xi_B \Gamma^{(1)}_{\lambda,B}(0,\phi) \right\}\, .
\ee
Now setting $n=1$ in \refb{efin} we get
\be
\delta
\Gamma^{(1)}_\lambda(0,\phi)
=  \Gamma^{(2)}_{\lambda}(0,\phi;0,\phi) 
\left\{ \delta\lambda + \sum_B \xi_B \Gamma^{(1)}_{\lambda,B}(0,\phi) \right\}\, .
\ee
$\delta\lambda$ has to be adjusted so that $\delta
\Gamma^{(1)}_\lambda(0,\phi)$ vanishes since $\Gamma^{(1)}_\lambda(0,\phi)$ 
vanishes both before and after the change.
This means that we must have
\be
\delta\lambda + \sum_B \xi_B \Gamma^{(1)}_{\lambda,B}(0,\phi)  = 0\, .
\ee
This in turn makes the right hand side of \refb{efin} vanish, showing that 
$\Gamma^{(n)}_\lambda(k_1,b_1;\cdots k_n,b_n)$ remains invariant under a change in the
local coordinate system.

Before concluding this section we would like to emphasize one important ingredient
underlying our analysis. In arriving at \refb{eshift3}, \refb{efeq} it was crucial that the
change in $\Lambda$ under a change of local coordinates had the form $\xi_A+\xi_B$
and not a quantity that has general dependence on both $A$ and $B$. This follows from the
fact that we used a set of gluing compatible local coordinates to define 1PI and 1PR amplitudes.
If instead we had chosen a more general local coordinate system -- {\it e.g.} one where the
choice of local coordinate on the puncture of the first Riemann surface depends on the
genus of the Riemann surface to which it is being glued -- and used it for
introducing the cut-off $\Lambda$ then this property will not be respected.

\sectiono{Extension to superstring and heterotic string theories} \label{super}

We shall now briefly discuss the extension of our analysis to superstring and heterotic
string theories when the field $\phi$ arises from the Neveu-Schwarz (NS) sector of heterotic
string theory or NSNS sector of type II string theory.
Let us for definiteness focus on the heterotic string  theory -- the generalization to the case of
superstrings is straightforward. The analysis proceeds more or less as in the case of bosonic
string theory; however
in this case 
the local coordinate system at a puncture requires specifying the holomorphic super-coordinates
$(w,\zeta)$ and anti-holomorphic coordinate $\wt w$. The generalization of \refb{eplumb} takes
the form\cite{1209.5461}
\be
w_1 w_2 = e^{-s+i\theta}, \quad w_2 \zeta_1 =\ve\, e^{-(s-i\theta)/2}\zeta_2,
\quad w_1\zeta_2 = -\ve\, e^{-(s-i\theta)/2} \, \zeta_1, \quad \zeta_1\zeta_2=0, \quad
 \wt w_1 \wt w_2 = e^{-s-i\theta}\, ,
\ee
with the $s$ integral running from 0 to $\Lambda$ in the definition of the propagator. 
$\ve$ takes value $\pm1$ and we have to sum over both values at the end to implement
GSO projection.
Now under a change of local 
(superconformal) coordinates $\Lambda$ still gets shifted by a term of the form $\xi_A+\xi_B$
as in \refb{eshift0}, but 
with $\xi_A$ possibly containing even nilpotent parts that depend on the super-moduli of the
punctured Riemann surface $A$ and $\xi_B$ containing even nilpotent parts that 
depend on the super-moduli of the
punctured Riemann surface $B$. As a result our analysis still goes through, with the sums over
$A$ and $B$ in various equations in \S\ref{s3} now being interpreted to include integration over
supermoduli space. 

\sectiono{Spurious infrared divergences} \label{sinfra}

Observations made below \refb{e6gen} show that the $n$-point truncated Green's function
$\Gamma^{(n)}_\lambda$ is equal to $\bar\Gamma^{(n)}_\lambda$. Since the latter is 
manifestly free from tadpole divergences, the $n$-point amplitude computed in the shifted
background is free from tadpole divergences. However the individual contributions to
$\bar\Gamma^{(n)}_\lambda$ can still suffer from spurious infrared divergence
from loop momentum integral since \refb{edefrule}
(with $\Gamma$ replaced by $\bar\Gamma$ on both sides) involves computing amplitudes with zero
momentum external $\phi$ legs in the original  vacuum. 
To understand the origin of these divergences and their resolution
it will be useful
to work with a finer
triangulation of the moduli space than the one discussed in \cite{1311.1257,1401.7014} for
dividing the contributions into 1PI and 1PR parts. This is done as follows:
\begin{enumerate}
\item We begin with a three punctured sphere with some specific 
choice of local coordinates at the
three punctures. The local coordinates are chosen to be symmetric under the exchange
of any two of the punctures.
\item Now we can generate a family of one punctured tori by gluing 
two of the punctures of the sphere by the propagator \refb{edefdelta}. We declare these 
one punctured tori to be composite one punctured tori. The rest of the one punctured tori 
are labelled as elementary. For the latter we can choose the local coordinate at the puncture in
an arbitrary fashion but it must match smoothly to those on the composite one punctured
tori across the codimension one subspace of the moduli space that divides composite
one-punctured tori from elementary one punctured tori.
\item Similarly by gluing two three punctured  spheres across one each of the punctures we can get
a family of four punctured spheres. We declare them to be composite four punctured spheres.
The rest of the four punctured spheres are declared as elementary. The choice of local coordinates
at the punctures of the latter
is arbitrary subject to the requirement of symmetry and smoothness across the
codimension one subspace of the moduli space that separates the elementary four punctured spheres
from composite four punctured spheres.
\item We now repeat the process iteratively. 
At the end we declare as composite 
all punctured Riemann surfaces which are
obtained by gluing two or more punctures of one or more 
elementary Riemann surfaces of lower genus and/or
lower number of
punctures by propagators. The rest of the Riemann surfaces are declared as elementary.
The full set of punctured Riemann surfaces contributing to an amplitude 
are then built in the same way that Feynman diagrams are built by gluing vertices with
propagators, with the elementary punctured
Riemann surfaces playing the role of 
the vertices. Indeed this is exactly the way the off-shell amplitudes are constructed in
string field theory in the Siegel gauge\cite{sft}, 
although we do not assume that the triangulation of the moduli space
we are using necessarily comes from any underlying gauge invariant string field theory.
\end{enumerate}

\begin{figure}
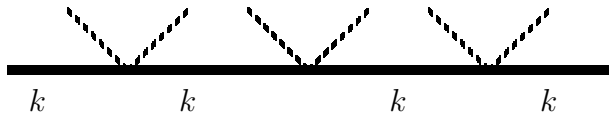

\begin{center}
\figg
\end{center}
\vskip -1.5 in
\caption{A subdiagram containing spurious infrared divergence. The thick line denotes 
a massless $\phi$ propagator carrying momentum $k$ and the dashed lines denote
$\lambda$ insertions.
\label{fg}}
\end{figure}

With this triangulation of the moduli space it is easy to see the origin of the spurious
infrared divergences. Consider
for example the case where we have an internal 
$\phi$ propagator carrying momentum $k$ propagating in the loop. Then by inserting a pair of
external zero momentum $\phi$ through the $\Gamma^{(4,0)}(k,\phi; -k,\phi;0,\phi;0,\phi)$ 
vertex we can increase the number
of internal 
$\phi$ propagators carrying the same momentum $k$ by 1. Repeating this process  
we can get a factor of (see Fig.~\ref{fg})
\be \label{emention}
(1/k^2)^n 
\ee
with arbitrary $n$ inside a diagram. For sufficiently large $n$ the integration over loop momentum
$k$ will give an infrared divergent contribution, invalidating perturbation theory.

If we use the language of quantum field theory with all the heavy fields integrated out, then
the solution to this problem is clear. Let $c$ denote the net contribution to 1PI 2-point function
of the originally masless fields
in the shifted background. Then the full propagator is given by
\be \label{efullphi}
{1\over k^2} + {1\over k^2}  c {1\over k^2}  + {1\over k^2}  c {1\over k^2}  c {1\over k^2}  +\cdots 
= {1\over k^2-c} \, .
\ee
Thus we can replace the massless propagator by \refb{efullphi} and drop all diagrams containing 
one or more insertions of 1PI 2-point function of massless fields. 
In this case there are no infrared divergences
arising from internal factors of the kind given in \refb{emention}.

The procedure described above requires computing the full 1PI two point function $c$ to a given order
for determining the modified propagator, but this is not necessary. Suppose $c$ has a power series
expansion in the coupling constant $g$ of the form $\sum_{n\ge 1}c_n g^{2n}\equiv c_1 g^2 + \delta c$.
Then it is sufficient to use the leading order correction $c_1 g^2$ to define the modified propagator
$1/(k^2 - c_1 g^2)$. The full propagator will now have an expansion of the form
\be \label{ebelow}
{1\over k^2 - c_1 g^2 - \delta c} = {1\over k^2 - c_1 g^2} + {1\over k^2 - c_1 g^2} \delta c 
{1\over k^2 - c_1 g^2} + {1\over k^2 - c_1 g^2} \delta c {1\over k^2 - c_1 g^2} 
\delta c {1\over k^2 - c_1 g^2}
+\cdots\, .
\ee
For $k^2 \sim g^2$ the factors of $1/(k^2 - c_1 g^2)$ can become of order $1/g^2$ but each such
factor will be accompanied by $\delta c \sim g^4$ and hence the successive terms in this expansion
will give smaller contributions.\footnote{The integration contour of $k$ can always be deformed 
in the complex plane to make $|k^2 - c_1 g^2|$ remain of order $g^2$ or larger. Note however 
that the $k$-integration will introduce 
non-analytic dependence on the shifted mass 
which in the present context will translate to non-analytic dependence
on the string coupling constant.}
Thus these terms can be treated perturbatively. 
In fact we can add an arbitrary order $g^4$ and higher 
contribution to $c_1 g^2$ in the definition of the modified propagator and subtract the
corresponding contribution from $\delta c$ without changing the final result.

In order to implement this procedure systematically in string theory we proceed 
as follows:
\begin{enumerate}
\item From the triangulation of the moduli space described above 
it is clear that possible infrared divergent contributions come only
from the composite Riemann surfaces where the integration variable $s$ 
of \refb{edefdelta} associated with one or more 
propagators becomes large. 
To isolate this divergence we split the propagator into its contribution
$\Delta_{\rm massless}$ from massless states\footnote{Here
massless states refer to fields which were massless in the original vacuum. For simplicity we
shall ignore the possibility of mixing between physical massless states with unphysical states
of the kind discussed in \cite{1401.7014}, but it should be possible to relax this assumption.}
 and the rest of the contribution $\wt\Delta$:
\be
\Delta = \Delta_{\rm massless} + \wt\Delta\, .
\ee
Note that unlike in the case of \refb{edefbardelta}, no cut-off $\Lambda$ on
$s$-integration is necesasry for definining $\Delta_{\rm massless}$ 
since we are working at non-zero momentum.
\item We also define $\wt\Gamma$ as the net contribution to an amplitude 
where all the $\Delta$'s
are replaced by $\wt\Delta$.
$\wt\Gamma$ defined this way is manifestly free from infrared divergences.
Furthermore the full amplitude can now be obtained by gluing $\wt\Gamma$'s by the propagator
$\Delta_{\rm massless}$. These are generically infrared 
divergent from the loop momentum integrals. Note that
diagrams where $\Delta_{\rm massless} $ connects a tadpole to the rest of the diagram vanish by our 
previous construction since all massless tadpoles have been made to cancel.
\item Let $C(k)$ denote the contribution to the two point function of massless fields
carrying momentum $k$ which are {\it 1PI in massless states, \i.e.\ do not contain an
internal $ \Delta_{\rm massless}$ propagator that is not part of a loop}. 
If there are more that one massless states then $C(k)$ is a matrix.
We now sum over all insertions of $C(k)$ into a propagator by using
\ben \label{eprop}
&& \Delta_{\rm massless} + \Delta_{\rm massless} C(k) \Delta_{\rm massless} +
\Delta_{\rm massless} C(k) \Delta_{\rm massless} C(k) \Delta_{\rm massless}+\cdots
\nonumber \\ &=& (\Delta_{\rm massless}^{-1} - C(k))^{-1}\, .
\een
The rule for computing the amplitudes to a given order in perturbation theory is now to replace
all internal $\Delta_{massless}$ factors by $(\Delta_{\rm massless}^{-1} - C(k))^{-1}$ with $C(k)$
computed to that particular order, and at the same time drop all contributions to the amplitude
that contain a $C(k)$ factor on an {\it internal leg}. 
This renders the amplitudes free from infrared divergence.
\item Note that the computation of $C(k)$ itself could suffer from infrared divergences
of the kind discussed above from subdiagrams. Thus the construction of $C(k)$ needs to
be carried out iteratively. 
We begin with the lowest order contribution to $C(k)$ which is free from infrared divergence
and define the lowest order modified propagator via \refb{eprop}. This is then used to
compute the next order contribution to $C(k)$ following the procedure described above.
This process is then repeated. In fact since the computation of $\lambda$ via 
\refb{eiter} also suffers from such spurious infrared divergences, this iterative procedure
must be carried out simultaneously for determining $\lambda$ and $C(k)$ to successively higher
order.
\item The expression for $C(k)$ obtained this way would in general depend on the
choice of local coordinates -- only the locations of the poles of the propagator in the $k^2$ plane
are free from ambiguity\cite{1311.1257,1401.7014}. 
This ambiguity essentially reflects the freedom of moving some contributions from the
propagators to the vertices and vice versa. However due to
the argument given below \refb{ebelow} we expect that the final result for any 
physical amplitude
should not suffer from any ambiguity as long as the procedure we are using renders them
free from any spurious infrared divergences.
\end{enumerate}

\bigskip

{\bf Acknowledgement:}
We thank Rajesh Gopakumar, Michael Green, Boris Pioline, Edward Witten 
and Barton Zwiebach for useful discussions.
The work of R.P. and A.S. was
supported in part by the 
DAE project 12-R\&D-HRI-5.02-0303. 
A.R. was supported by the Ramanujan studentship of Trinity College, 
Cambridge, and his research leading to these results 
has also 
received funding from the European Research Council under 
the European Community's Seventh Framework Programme 
(FP7/2007-2013) / ERC grant agreement no. [247252].
The work of A.S. was also supported in
part by the
J. C. Bose fellowship of 
the Department of Science and Technology, India.

\appendix

\sectiono{Effect of shifting a massless field} \label{sa}

Let us consider a quantum field theory containing a 
massless field $\phi$ and a set of other massless
and massive fields. 
Let us suppose that we have computed 
the 1PI amplitude involving these fields. Then the truncated Green's functions
can be computed by summing up the contributions from {\it tree level} graphs with the
propagators and vertices constructed from the 1PI amplitudes. 
We shall study how these truncated Green's functions change when we shift
the background value of the field
$\phi$ by a constant $\lambda$,
without necessarily assuming that the background is a solution
to the equations of motion. Our goal will be to prove \refb{edefrule}.

Now since eventually we shall be interested in setting the background value of $\phi$ to
a solution to its equations of motion, or equivalently demand that the tadpole of the
field $\phi$ vanishes, it will be natural to demand that all other fields also
satisfy their equations of motion. Does this require shifting other fields as well?
As in the text, we shall assume that none of the other massless
fields need to be shifted even when $\phi$ is shifted, so we only have to worry about a possible
shift of the massive fields. To this end we note that only zero momentum modes of the massive
fields may be shifted. Since our goal will be to analyze amplitudes where all the external states
carry generic non-zero momentum, except possibly some zero momentum $\phi$ fields, we
shall integrate out all the zero momentum modes of massive fields, \i.e.\ include in the
1PI amplitude also those which are one particle reducible in one or more zero momentum 
propagator of massive states. Now the only shift will be in the $\phi$ field, and eventually
when we require $\phi$ to satisfy its equation of motion all the massive fields will automatically
satisfy their equations of motion.

To proceed we note that \refb{edefrule} is equivalent to
\be \label{equiv}
{\p \Gamma^{(n)}_\lambda (k_1, b_1; \cdots k_n,b_n)\over \p\lambda}
=  \Gamma^{(n+1)}_\lambda (k_1, b_1; \cdots k_n,b_n; 0, \phi)\, .
\ee
Thus it is enough to establish \refb{equiv} for arbitrary $\lambda$.
To compute the left hand side we shall divide the 1PI action into its kinetic term and the
interaction term, and include in the kinetic term only genuine tree level contributions
at $\lambda=0$,
including the rest of the contribution into the interaction term. Then neither the kinetic
terms, nor the $\prod_i (k_i^2 + m_{b_i}^2)$ terms appearing in \refb{eoff}, have any
dependence on $\lambda$. 
Let $\wt\psi_{b}(k)$ denote the general set of fields in the momentum space. 
If $V_\lambda$ denotes the interaction term in the 1PI action in the shifted
background,
then, by noting that $\phi$ and $\lambda$ must appear in the combination $\phi+\lambda$
in the interaction part of the 1PI action,
we see that 
\ben \label{eleft}
{\p \Gamma^{(n)}_\lambda (k_1, b_1; \cdots k_n,b_n)\over \p\lambda}
&=& \prod_{i=1}^n (k_i^2 + m_{b_i}^2) \left\langle 
\prod_{i=1}^n \wt\psi_{b_i}(k_i) \left(-{\p V_\lambda\over \p\lambda}\right) \, 
\right\rangle \nonumber \\
&=& \prod_{i=1}^n (k_i^2 + m_{b_i}^2) \left\langle 
\prod_{i=1}^n \wt\psi_{b_i}(k_i) \left(-{\delta V_\lambda\over \delta \wt\phi(0)}\right) \, 
\right\rangle\, ,
\een
where $\langle~\rangle$ denotes the tree level amplitude computed with the 1PI
action.
Let us now examine the right hand side of \refb{equiv}. This has to be interpreted 
as the $k\to 0$ limit of the expression where $0,\phi$ is replaced by $k,\phi$.
Since the genuine tree level mass of $\phi$ is zero, we have
\be \label{eright}
 \Gamma^{(n+1)}_\lambda (k_1, b_1; \cdots k_n,b_n; k, \phi) =
  \prod_{i=1}^n (k_i^2 + m_{b_i}^2) \left\langle 
\prod_{i=1}^n \wt\psi_{b_i}(k_i)  (k^2 \,\wt \phi(k)) \, 
\right\rangle\, .
\ee
Now the equation of motion of $\phi$ computed from the 1PI action is
\be
k^2 \, \wt\phi(k) + \left({\delta V_\lambda\over \delta \wt\phi(-k)}\right) =0\, .
\ee
Since equation of motion inserted into tree level amplitude computed from 1PI action
vanishes, we get
\be 
\prod_{i=1}^n (k_i^2 + m_{b_i}^2) \left\langle 
\prod_{i=1}^n \wt\psi_{b_i}(k_i)  (k^2 \,\wt \phi(k))\right\rangle
= \prod_{i=1}^n (k_i^2 + m_{b_i}^2) \left\langle 
\prod_{i=1}^n \wt\psi_{b_i}(k_i)  \left(-{\delta V_\lambda\over \delta \wt\phi(-k)}\right)
\right\rangle\, .
\ee
Taking the $k\to 0$ limit of this expression we establish the equivalence of
\refb{eleft} and \refb{eright}. This in turn proves \refb{equiv}.

We would like to end with the remark 
that the individual Feynman diagrams contributing to
both sides of \refb{equiv} (and \refb{edefrule}) are divergent as they may contain zero
momentum internal $\phi$-propagators. Thus at this stage our analysis should be taken as
a proof of equality of the combinatorial factors which appear when we express the two sides
of \refb{edefrule} as a sum of Feynman diagrams in the $\lambda=0$ theory. Once the
infrared divergences on both sides of  \refb{edefrule} are
regularized using \refb{edeltareg}, it becomes a true
algebraic equality.




\begin{thebibliography}{99}

\bibitem{DSW}
 M.~Dine, N.~Seiberg and E.~Witten,
  Nucl.\ Phys.\ B {\bf 289}, 589 (1987).

\bibitem{ADS}
J.~J.~Atick, L.~J.~Dixon and A.~Sen,
  ``String Calculation of Fayet-Iliopoulos d Terms in Arbitrary Supersymmetric Compactifications,''
  Nucl.\ Phys.\ B {\bf 292}, 109 (1987).

\bibitem{DIS}
M.~Dine, I.~Ichinose and N.~Seiberg,
  ``F Terms and d Terms in String Theory,''
  Nucl.\ Phys.\ B {\bf 293}, 253 (1987).
  
\bibitem{greenseiberg} 
  M.~B.~Green and N.~Seiberg,
  ``Contact Interactions in Superstring Theory,''
  Nucl.\ Phys.\ B {\bf 299}, 559 (1988).

\bibitem{1304.2832}
E.~Witten,
``More On Superstring Perturbation Theory,''
  arXiv:1304.2832 [hep-th].
  
 \bibitem{1404.5346}
 N.~Berkovits and E.~Witten,
``Supersymmetry Breaking Effects
using the Pure Spinor Formalism of the Superstring",  
arXiv:1404.5346 [hep-th].


\bibitem{1209.5461} 
  E.~Witten,
  ``Superstring Perturbation Theory Revisited,''
  arXiv:1209.5461 [hep-th].

\bibitem{Belopolsky} 
  A.~Belopolsky,
  ``De Rham cohomology of the supermanifolds and superstring BRST cohomology,''
  Phys.\ Lett.\ B {\bf 403}, 47 (1997)
  [hep-th/9609220];
``New geometrical approach to superstrings,''
  hep-th/9703183;
``Picture changing operators in supergeometry and superstring theory,''
  hep-th/9706033.

\bibitem{dp}
  E.~D'Hoker and D.~H.~Phong,
  ``Two loop superstrings. I. Main formulas,''
  Phys.\ Lett.\ B {\bf 529}, 241 (2002)
  [hep-th/0110247].
``II. The Chiral measure on moduli space,''
  Nucl.\ Phys.\ B {\bf 636}, 3 (2002)
  [hep-th/0110283].
``III. Slice independence and absence of ambiguities,''
  Nucl.\ Phys.\ B {\bf 636}, 61 (2002)
  [hep-th/0111016].
``IV: The Cosmological constant and modular forms,''
  Nucl.\ Phys.\ B {\bf 639}, 129 (2002)
  [hep-th/0111040].
``V. Gauge slice independence of the N-point function,''
  Nucl.\ Phys.\ B {\bf 715}, 91 (2005)
  [hep-th/0501196].
``VI: Non-renormalization theorems and the 4-point function,''
  Nucl.\ Phys.\ B {\bf 715}, 3 (2005)
  [hep-th/0501197].
``VII. Cohomology of Chiral Amplitudes,''
  Nucl.\ Phys.\ B {\bf 804}, 421 (2008)
  [arXiv:0711.4314 [hep-th]].


\bibitem{Witten}
E.~Witten,
  ``Notes On Supermanifolds and Integration,''
  arXiv:1209.2199 [hep-th];
 ``Notes On Super Riemann Surfaces And Their Moduli,''
  arXiv:1209.2459 [hep-th];
``Notes On Holomorphic String And Superstring Theory Measures Of Low Genus,''
  arXiv:1306.3621 [hep-th];
``The Feynman $i \epsilon$ in String Theory,''
  arXiv:1307.5124 [hep-th].

\bibitem{1304.7798} 
  R.~Donagi and E.~Witten,
  ``Supermoduli Space Is Not Projected,''
  arXiv:1304.7798 [hep-th].

\bibitem{sft}
B.~Zwiebach,
  ``Closed string field theory: Quantum action and the B-V master equation,''
  Nucl.\ Phys.\ B {\bf 390}, 33 (1993)
  [hep-th/9206084].

\bibitem{FS} 
  W.~Fischler and L.~Susskind,
  ``Dilaton Tadpoles, String Condensates and Scale Invariance,''
  Phys.\ Lett.\ B {\bf 171}, 383 (1986).
  ``Dilaton Tadpoles, String Condensates and Scale Invariance. 2.,''
  Phys.\ Lett.\ B {\bf 173}, 262 (1986).

\bibitem{earlyref} 
   B.~W.~Lee,
  ``Renormalization of the sigma model,''
  Nucl.\ Phys.\ B {\bf 9}, 649 (1969).
  K.~Bardakci,
  ``Dual models and spontaneous symmetry breaking,''
  Nucl.\ Phys.\ B {\bf 68}, 331 (1974);
``Dual models and spontaneous symmetry breaking ii,''
  Nucl.\ Phys.\ B {\bf 70}, 397 (1974);
  K.~Bardakci and M.~B.~Halpern,
  ``Explicit Spontaneous Breakdown in a Dual Model,''
  Phys.\ Rev.\ D {\bf 10}, 4230 (1974).

\bibitem{0410101} 
  E.~Dudas, G.~Pradisi, M.~Nicolosi and A.~Sagnotti,
  ``On tadpoles and vacuum redefinitions in string theory,''
  Nucl.\ Phys.\ B {\bf 708}, 3 (2005)
  [hep-th/0410101].

\bibitem{AtickS}
  J.~J.~Atick and A.~Sen,
  ``Two Loop Dilaton Tadpole Induced by Fayet-iliopoulos $D$ Terms in Compactified Heterotic String Theories,''
  Nucl.\ Phys.\ B {\bf 296}, 157 (1988).


  \bibitem{1403.5494} 
  E.~D'Hoker and D.~H.~Phong,
  ``Two-loop vacuum energy for Calabi-Yau orbifold models,''
  Nucl.\ Phys.\ B {\bf 877}, 343 (2013)
  [arXiv:1307.1749];
  E.~D'Hoker 
  Topics in Two-Loop Superstring Perturbation Theory
arXiv:1403.5494 [hep-th];



\bibitem{1311.1257} 
  R.~Pius, A.~Rudra and A.~Sen,
  ``Mass Renormalization in String Theory: Special States,''
  arXiv:1311.1257 [hep-th].

\bibitem{1401.7014} 
  R.~Pius, A.~Rudra and A.~Sen,
  ``Mass Renormalization in String Theory: General States,''
  arXiv:1401.7014 [hep-th].

\bibitem{catoptric} 
  J.~J.~Atick, G.~W.~Moore and A.~Sen,
  ``Catoptric Tadpoles,''
  Nucl.\ Phys.\ B {\bf 307}, 221 (1988).

\end{thebibliography}
\end{document}